\documentclass[final,5p,times,twocolumn]{elsarticle}

\usepackage[utf8]{inputenc}
\usepackage{amsmath}
\usepackage{graphicx}
\usepackage{siunitx}
\DeclareSIUnit\angstrom{\text{\AA}}
\usepackage{textcomp}
\usepackage{gensymb}
\usepackage{microtype}
\usepackage{amsfonts}
\usepackage{lmodern}
\usepackage{caption}
\usepackage[colorlinks=true,linkcolor=blue,citecolor=blue,urlcolor=blue]{hyperref}

\makeatletter
\def\ps@pprintTitle{%
  \let\@oddhead\@empty
  \let\@evenhead\@empty
  \def\@oddfoot{}%
  \let\@evenfoot\@oddfoot}
\makeatother

\begin{document}

\begin{frontmatter}

  \title{First Plasma Atomic Layer Etching of Diamond via O$_2$/Kr Chemistry}

  \author[insta,instb,instc]{D.D. Tran\corref{cor1}}
  \author[instc,instd]{C. Mannequin}
  \author[instb,instc]{A. Traore}
  \author[instb]{M. Sasaki}
  \author[insta,instb,instc]{E. Gheeraert}

  \cortext[cor1]{Corresponding author}

  \address[insta]{Univ. Grenoble Alpes, CNRS, Grenoble INP, Institut Neel, Grenoble 38000, France}
  \address[instb]{Institute of Applied Physics, Faculty of Pure and Applied Sciences, University of Tsukuba, Tsukuba 305-8573, Japan}
  \address[instc]{Japanese-French laboratory for Semiconductor Physics and Technology J-FAST, CNRS, Université Grenoble Alpes, Grenoble INP, University of Tsukuba, Japan}
  \address[instd]{Nantes Université, CNRS, Institut des Matériaux de Nantes Jean Rouxel, IMN, F-44000 Nantes, France}

  \begin{abstract}

   We report the first plasma atomic layer etching (ALE) process for diamond using a cyclic plasma sequence composed of two separated steps: oxygen surface modification and krypton ion removal. The process is implemented in an inductively coupled plasma reactor using alternating O$_2$ plasma exposure and low-energy Kr ion bombardment.

    This cyclic process exhibits the characteristic self-limiting behavior of ALE and enables controlled material removal with atomic-scale precision. An etch depth per cycle of \SI{6.85}{\angstrom} was achieved. Surface analysis reveals that the etched diamond surfaces exhibit lower roughness than the pristine material, while XPS confirms the preservation of the diamond bonding structure and indicates essentially damage-free etching.

    These results demonstrate that plasma ALE based on O$_2$/Kr chemistry provides a viable route toward damage-controlled nanoscale processing of diamond, opening new opportunities for advanced device fabrication in power electronics, photonics, quantum sensing and quantum computing technologies.

  \end{abstract}


\end{frontmatter}

\section{Introduction}

Diamond has emerged as a promising semiconductor material owing to its exceptional physical properties, including an ultra-wide bandgap (\SI{5.5}{eV}) \cite{umezawa2018recent}, high carrier mobility (up to $10^6$ cm$^2$·V$^{-1}$·s$^{-1}$) \cite{portier2023carrier}, a high breakdown field (\SI{9.5}{MV.cm^{-1}}) \cite{volpe2010high}, and the highest known thermal conductivity (\SI{22}{W.cm^{-1}.K^{-1}} at room temperature) \cite{umezawa2018recent}. These characteristics make diamond highly attractive for applications in high-power electronics and photonic devices \cite{kah2025diamond, canas2021normally}. Moreover, the presence of nitrogen-vacancy (NV) centers makes diamond a promising platform for quantum sensing and quantum computing technologies \cite{pezzagna2021quantum}.

Despite these advantages, the fabrication of diamond devices remains challenging, largely because of the difficulty of achieving precise and low-damage etching. The strong sp$^3$ carbon--carbon bonds responsible for diamond's remarkable mechanical and chemical stability also make it highly resistant to conventional plasma etching processes.

As a result, current diamond etching techniques such as ion beam etching (IBE) and reactive ion etching (RIE) generally require relatively high ion energies to remove material \cite{toros2020reactive}. Although these approaches can provide reasonable etch rates, they often induce surface roughness, subsurface structure damage, and partial conversion of the diamond surface toward graphitic phases \cite{kawabata2004xps}, creating photonic dead layers. Such defects can degrade device performance, particularly in applications requiring high-quality surfaces, sharp interfaces, or nanoscale control of etch depth. Reducing the density of diamond surface states by approximately threefold can increase MOSFET field-effect mobility by up to hundred times \cite{matsumoto2019inversion}. 

Atomic layer etching is an ideal approach for achieving atomic-scale precision and damage-free in material processing. ALE relies on cyclic sequences of two separated steps: a surface-modification followed by a removal. Depending on the target material, surface modification can be achieved through chemisorption, deposition, conversion, or extraction, whereas the removal step is typically accomplished by physical ion bombardment or thermal activation \cite{lill2021atomic}.

ALE is governed by two key criteria: separation of the modification and removal half-reactions, and self-limiting behavior of at least one step, and ideally both steps. Together, these conditions ensure that, in each cycle, only a finite number of surface monolayers are modified and subsequently removed, leading to atomic-scale control of etch per cycle.

\begin{figure}[t]
  \centering
  \includegraphics[width=\columnwidth]{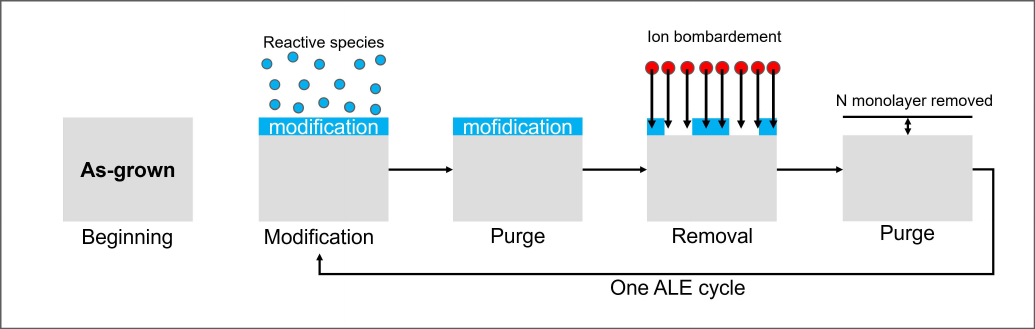}
  \caption{Schematic illustration of the plasma atomic layer etching cycle used in this work, consisting of surface modification by reactive species followed by ion removal, with purge steps inserted between the two half-cycles.}
  \label{fig:ale_cycle}
\end{figure}

Although ALE has been demonstrated for several semiconductors \cite{kanarik2015overview,carver2015atomic,oehrlein2015atomic,mannequin2020comparative}, plasma ALE of diamond has not yet been experimentally reported. A concept for diamond ALE was proposed by Yoder in 1988 \cite{Yoder1988}, involving NO$_2$ surface modification followed by ion-induced removal. However, this approach was not experimentally demonstrated, and the underlying mechanism remained unclear.

In this study, we investigate plasma ALE, where modification occurs via chemisorption of reactive plasma species and removal is carried out by ion bombardment, as illustrated in \autoref{fig:ale_cycle}. A central challenge in diamond ALE is the high strength of C--C bonds. For an ideal ALE process, the carbon--adsorbate bond should be stronger than the carbon--carbon bond so that chemical modification can effectively destabilize the topmost surface layer for selective removal. For most plasma chemistries, bond-energy comparisons for diatomic carbon bonds indicate that only C--O and C--N bonds are stronger than the surface C--C bond, which makes oxygen- and nitrogen-based chemistries the most suitable candidates for the modification step. Among these options, oxygen is particularly favorable because the C--O bond energy is nearly twice that of the C--C bond \cite{haynes2016crc} and the diffusion of oxygen is limited on diamond surface, enabling efficient surface modification while maintaining a controllable ALE pathway. For the removal step, heavy noble gas ions such as Kr offer directional and controllable ion bombardment. Establishing a practical ALE regime therefore requires careful investigation of both surface modification and selective removal.

In this work, we demonstrate the first plasma ALE process for diamond using oxygen plasma for surface modification and krypton ion bombardment for removal. The cyclic O$_2$/Kr process is investigated in terms of the self-limiting ALE window, reaction separation - process synergy, removal-step dynamics, and etching performance. The results demonstrate controlled diamond etching with sub-nanometer precision, essential damage-free surface and subsurface, and provide insight into the mechanism governing the ALE process.

\section{Experimental Methods}

Experiments were performed using a Samco inductively coupled plasma (ICP) reactor configured for cyclic plasma processing. The ALE process consists of repeated cycles composed of two sequential steps:

\textbf{Surface modification step.}
An oxygen plasma is applied to modify the diamond surface, producing oxygen-terminated carbon bonds and oxidized surface species.

\textbf{Removal step.}
Low-energy krypton ions are then used to remove the modified surface layer through directional ion bombardment.

\begin{figure}[t]
  \centering
  \includegraphics[width=\columnwidth]{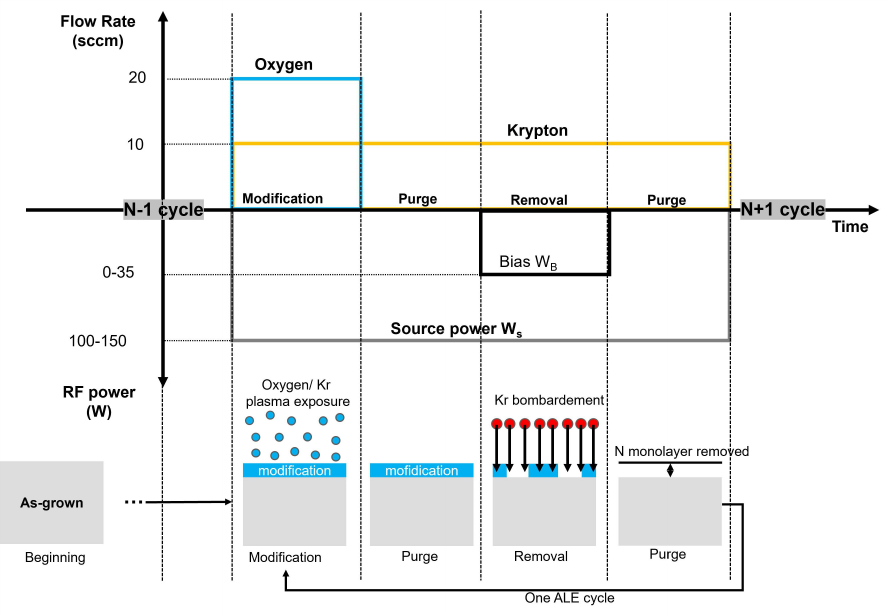}
  \caption{One ALE cycle consists of four sequential steps. Krypton is present throughout the cycle, whereas oxygen is introduced only during the surface-modification step. The source power is kept constant, and RF bias is applied only during the removal step.}
  \label{fig:ale_exp_cycle}
\end{figure}

Between each step, the chamber is purged to prevent overlap between the chemical modification and physical removal stages, thereby maintaining the separation of the two half-reactions characteristic of ALE. The detailed four-step timing sequence used in this work is shown in \autoref{fig:ale_exp_cycle}.

\begin{figure}[t]
  \centering
  \includegraphics[width=\columnwidth]{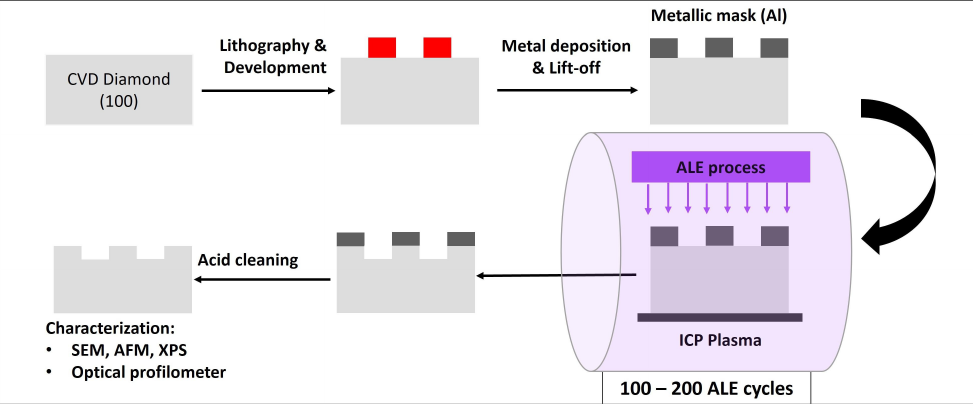}
  \caption{ALE experiment flow. After cleaning, lithography, and aluminum mask patterning, the diamond sample is etched inside an ICP reactor. After etching, the aluminum mask is removed using an aluminum etching solution. The etched diamond surface is then characterized by SEM, optical profilometry, AFM, and XPS.}
  \label{fig:ale_process}
\end{figure}

Single-crystal CVD diamond substrates were patterned with an aluminum mask using standard microfabrication procedures, including lithographic patterning, thin-film ion-sputtering deposition, and lift-off, as shown in \autoref{fig:ale_process}. After the ALE process, the aluminum mask was removed using Transene A aluminum etchant, consisting of 55--65 wt.\% phosphoric acid, 1--5 wt.\% nitric acid, and 3--5 wt.\% acetic acid, at \SI{60}{\celsius} for 10 minutes. The sample was then cleaned sequentially with deionized water, isopropanol, and acetone.

Surface morphology and surface chemistry were characterized using scanning electron microscopy (SEM), atomic force microscopy (AFM), and X-ray photoelectron spectroscopy (XPS). The etch depth was determined by optical profilometry. The etch per cycle (EPC) was calculated by dividing the final etch depth by the number of ALE cycles.

\section{Results and Discussion}

\subsection{ALE Window}

\begin{figure}[t]
  \centering
  \includegraphics[width=\columnwidth]{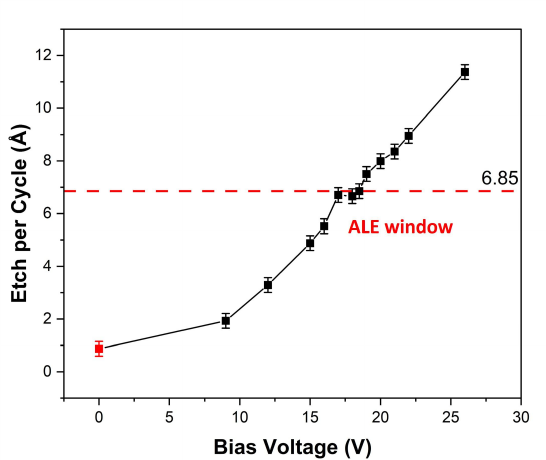}
  \caption{Etch per cycle (EPC) as a function of removal-step DC bias for the O$_2$/Kr ALE process. The trend identifies the ion-energy regime where the surface-modified diamond is preferentially removed, defining the practical ALE operating window. The red dot corresponds to etching without the removal step.}
  \label{fig:ale_window_energy_scan}
\end{figure}

To identify the self-limiting regime of the process, the etch per cycle was investigated as a function of ion energy during the removal step (\autoref{fig:ale_window_energy_scan}). In an ICP reactor, the incident ion energy at the substrate can be approximated as the sum of the plasma potential ($V_p$) and the applied DC bias ($|V_{\mathrm{bias}}|$), such that $E_{\mathrm{ion}} \approx e(V_p + |V_{\mathrm{bias}}|)$, where $V_p$ is typically on the order of \SIrange{20}{30}{\volt}.

During the modification step, oxygen radicals chemisorb on the diamond surface, forming oxidized carbon species such as C--O and C=O and weakening the surface C--C bonds. This chemical modification reduces the effective sputtering threshold of the topmost layer relative to that of the underlying bulk diamond. The difference in sputtering thresholds creates a finite ion-energy window in which only the modified layer is removed, while the underlying lattice remains largely unaffected. In this regime, the EPC becomes nearly independent of ion energy, which is characteristic of self-limiting ALE behavior.

The EPC as a function of removal-step bias reveals a narrow ALE window of approximately \SI{1.5}{eV}, corresponding to a bias voltage range of about \SIrange{17}{18.5}{V} and a EPC of \SI{6.85}{\angstrom}. Assuming a plasma potential of approximately \SI{20}{V}, the corresponding ion energy is estimated to be in the range of \SIrange{37}{38.5}{eV}. Within this window, the ion energy is sufficient to remove the oxidized surface layer while remaining below the sputtering threshold of bulk diamond.

The narrow ALE window indicates that the difference between the sputtering thresholds of the oxygen-modified surface layer and pristine diamond is small. As a result, precise control of the ion energy is required to achieve selective and self-limiting etching. Despite its limited width (~1.5 eV), the transition between the ALE and sputtering regimes is clearly resolved and exceeds the experimental uncertainty.

\subsection{Process Synergy}

\begin{figure}[t]
  \centering
  \includegraphics[width=\columnwidth]{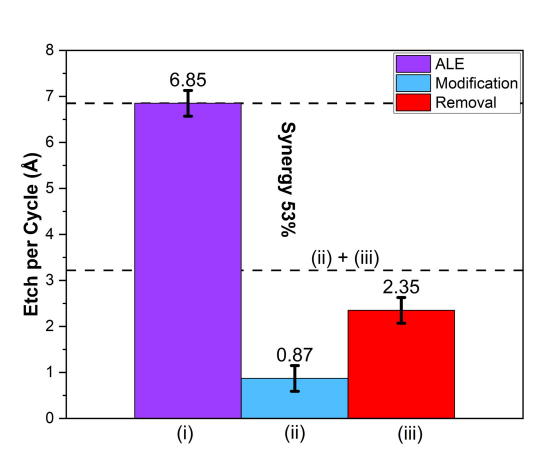}
  \caption{Quantitative process-synergy analysis of the O$_2$/Kr diamond ALE cycle. Bar (i) shows the etch per cycle (EPC) for the full ALE sequence, while bars (ii) and (iii) show the EPC from modification-only and removal-only control cycles, respectively. The ALE EPC (\SI{6.85}{\angstrom}) exceeds the sum of the two individual contributions (\SI{0.87}{\angstrom} + \SI{2.35}{\angstrom}), yielding a synergy of approximately 53\%.}
  \label{fig:ale_synergy}
\end{figure}

To evaluate the separation of the modification and removal half-reactions, control experiments were performed using cyclic processes containing only the modification step or only the removal step (\autoref{fig:ale_synergy}).

When the oxygen modification step is applied alone, a very small but non-zero EPC of \SI{0.87}{\angstrom} is observed. This behavior is attributed to weak ion bombardment during the modification step, since ions in the plasma still acquire a small acceleration energy, typically on the order of 5--10 eV, because of the difference between the plasma potential and the floating potential of the surface. Over many cycles, this weak bombardment can induce slow reactive ion etching.

Similarly, a cyclic process containing only the krypton removal step also produces a non-zero EPC of \SI{2.35}{\angstrom} due to progressive sputtering of the diamond surface during repeated low-energy ion bombardment. This behavior may be associated with a fatigue effect, in which cumulative ion impacts gradually lower the sputtering threshold of the surface.

When both steps are applied sequentially in a complete ALE cycle, the measured etch per cycle is significantly larger than that obtained in either half-cycle alone. The calculated synergy for the O$_2$/Kr process is approximately 53\%, confirming that the dominant material removal arises from the cyclic interaction between surface oxidation and ion-assisted removal rather than from the isolated action of either step.

\subsection{Removal-Step Time}

\begin{figure}[t]
  \centering
  \includegraphics[width=\columnwidth]{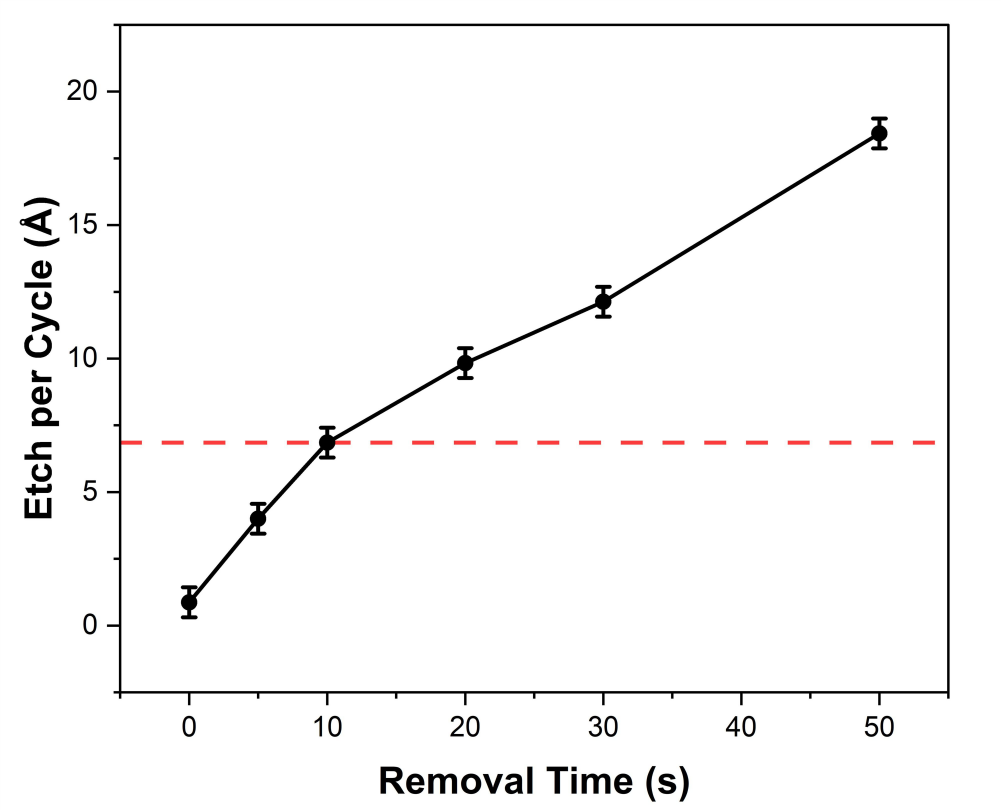}
  \caption{Etch per cycle as a function of krypton ion removal time for the O$_2$/Kr ALE process. EPC increases with removal time, indicating progressive removal of the oxygen-modified surface layer. The deviation from a fully saturated plateau suggests that the removal step is not perfectly self-limiting under these conditions, with extended ion exposure leading to additional interaction with the underlying diamond.}
  \label{fig:epc_vs_t_activation}
\end{figure}

The influence of the krypton ion removal step time was investigated by varying the ion-bombardment duration while maintaining constant plasma conditions. The evolution of the EPC as a function of removal time provides insight into the kinetics of ion-assisted removal, as shown in \autoref{fig:epc_vs_t_activation}.

As the removal time increases, the EPC initially rises rapidly, indicating efficient removal of the oxygen-modified surface layer formed during the preceding modification step. In this regime, Kr ion bombardment promotes the desorption of oxidized carbon species, resulting in progressive removal of the modified layer.

Beyond a characteristic removal duration of \SI{10}{s}, the increase in EPC becomes significantly slower. This transition indicates that most of the modified layer has already been removed and that additional ion bombardment begins to interact with the underlying diamond lattice.

Unlike an ideal ALE process, in which complete saturation would occur once the modified layer is fully removed, a gradual increase in EPC persists at longer removal times. This behavior can be attributed to limited sputtering of bulk diamond, caused by the presence of higher-energy ions in the ion energy distribution, as well as cumulative ion-induced surface damage that progressively lowers the bulk sputtering threshold. This effect will be discussed in more detail in a subsequent article comparing the use of different noble gases in the ALE removal step.

\subsection{Surface Morphology and Roughness}
Surface morphology and roughness were analyzed to evaluate etch-induced damage and smoothing effects.

\begin{figure}[t]
  \centering
  \includegraphics[width=\columnwidth]{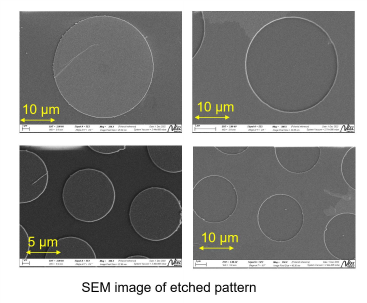}
  \caption{SEM images of ALE-patterned diamond structures, showing smooth etched surfaces without obvious micromasking or edge damage. The patterns on the left correspond to cylindrical islands, whereas the patterns on the right exhibit hollow circular geometries.}
  \label{fig:sem_surface_quality}
\end{figure}

The etched structures were first analyzed by SEM, as shown in \autoref{fig:sem_surface_quality}. Patterned diamond features fabricated using the ALE process exhibit well-defined vertical sidewalls and uniform etching profiles across the patterned regions. The etched surfaces appear smooth and free of significant micromasking or edge damage, indicating that the ALE process maintains good anisotropy while preserving surface quality.

\begin{figure}[t]
  \centering
  \includegraphics[width=0.49\columnwidth]{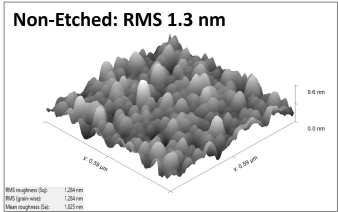}\hfill
  \includegraphics[width=0.49\columnwidth]{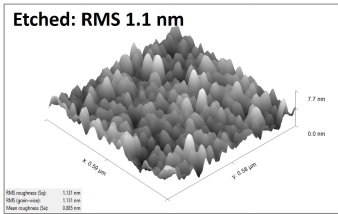}
  \caption{Surface topographies of diamond measured by AFM: (left) pristine (non-etched) diamond surface with an RMS roughness of \SI{1.23}{nm}, and (right) diamond surface etched within the ALE window at a removal bias of \SI{18}{V}, with an RMS roughness of \SI{1.1}{nm}.}
  \label{fig:afm_before_after}
\end{figure}

Surface morphology was then examined by atomic force microscopy, as shown in \autoref{fig:afm_before_after} for diamond etched within the ALE window at a removal bias of \SI{18}{V}. The root-mean-square (RMS) roughness decreases from approximately \SI{1.23}{nm} for the pristine diamond surface to about \SI{1.1}{nm} for the etched surface. This trend indicates that etching within the ALE window removes only a few monolayers per cycle while preserving the original surface morphology.

In addition, repeated low-energy removal cycles can gradually eliminate pre-existing surface defects through gentle sputtering. The ALE process therefore not only avoids significant surface degradation but can also provide slight surface smoothing during cyclic processing.

\subsection{Surface Chemistry}

\begin{figure}[t]
  \centering
  \includegraphics[width=\columnwidth]{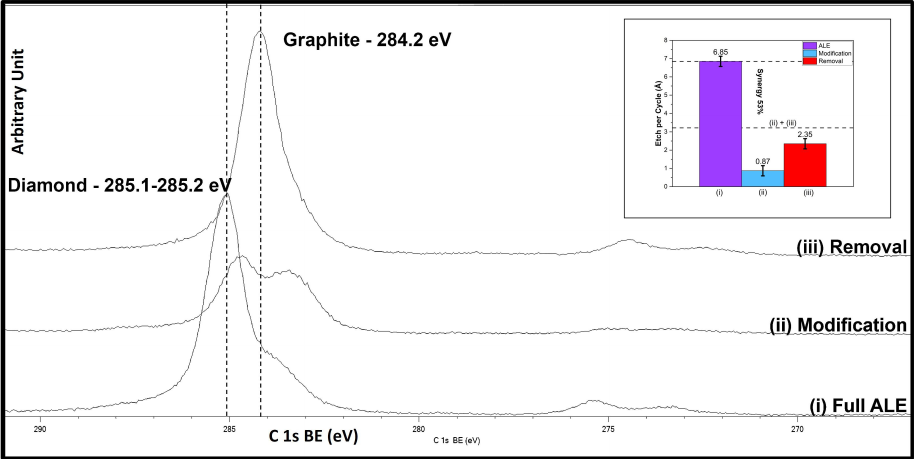}
  \caption{C~1s XPS spectra of diamond after three processing conditions: (i) full O$_2$/Kr ALE at a removal bias of \SI{18}{V}, (ii) cyclic modification-only process (oxygen plasma step only), and (iii) cyclic removal-only process (krypton ion step only). The full ALE condition preserves a dominant diamond-related C--C component near 285.1--285.2~eV with only a weak graphite-like contribution near 284.2~eV, indicating limited graphitization compared with the single-step control processes.}
  \label{fig:xps_c1s}
\end{figure}

Finally, the surface chemical state was analyzed by X-ray photoelectron spectroscopy (XPS), as shown in \autoref{fig:xps_c1s}. The C~1s spectra were compared for three conditions from the synergy test: (i) the complete ALE cycle performed at a removal bias of \SI{18}{V}, (ii) the modification-only process, and (iii) the removal-only process.

For the removal-only process, the C~1s spectrum exhibits a peak near 284.2~eV, indicating damage to the diamond structure and the formation of graphitic or disordered carbon. This behavior is attributed to repeated low-energy Kr ion bombardment without the protective effect of cyclic surface modification, similar to ion-beam-etching damage reported in \cite{kawabata2004xps}.

For the modification-only process, the spectrum shows several contributions between 284 and 285~eV, also indicating progressive damage and partial graphitization of the diamond surface, although less pronounced than in the removal-only case. This XPS spectrum is close to that reported for RIE etching of diamond in \cite{kawabata2004xps}.

A weak feature between 283 and 284~eV is observed in all spectra and is attributed to a SiC-like contribution arising from cross-contamination from the Si carrier wafer.

In contrast, for the full ALE process, the spectrum is dominated by the peak at 285.1--285.2~eV, characteristic of sp$^3$-bonded diamond, while only a very weak contribution is observed near 284.2~eV. This indicates that the cyclic alternation of surface modification and ion-assisted removal enables selective removal of the modified layer while largely preserving the underlying diamond structure. These results highlight the advantage of the ALE sequence over the individual half-cycles, as it minimizes surface and subsurface damage compared with conventional etching behavior.

Taken together, these results demonstrate that the O$_2$/Kr ALE process enables controlled layer-by-layer removal of diamond while largely preserving both the chemical bonding structure and the surface morphology. This work therefore represents the first experimental demonstration of plasma atomic layer etching of diamond.

The combination of atomic-scale etch control, preservation of the diamond bonding structure, and improved surface morphology demonstrates the effectiveness of the O$_2$/Kr ALE process for precise diamond nanofabrication. These characteristics are particularly important for diamond-based electronic, photonic and quantum devices, where structure damage and roughness strongly influence device performance.

\section{Mechanism of Diamond Plasma ALE}

The O$_2$/Kr plasma atomic layer etching process developed in this work is governed by the cyclic interaction between oxygen plasma surface modification and low-energy ion-assisted removal.

During the oxygen plasma step, reactive oxygen species chemisorb on the diamond surface and form oxygen-containing functional groups. This surface oxidation weakens the surface C--C bonds and lowers the sputtering threshold of the modified layer compared with that of the underlying diamond lattice.

In the subsequent removal step, the surface is exposed to low-energy Kr ion bombardment. When the ion energy is tuned within the ALE window, it is sufficient to remove the oxidized surface layer while remaining below the sputtering threshold of bulk diamond. The ion bombardment breaks the weakened surface bonds and promotes the desorption of oxidized carbon species, resulting in removal of the modified surface layer.

Because the sputtering threshold of the modified layer is lower than that of the underlying diamond, ion-assisted removal primarily affects the oxidized surface region. Once this modified layer has been removed, further sputtering becomes energetically less favorable, leading to the self-limiting behavior characteristic of atomic layer etching.

The cyclic repetition of these two steps therefore enables controlled layer-by-layer removal of diamond with atomic-scale precision. Deviations from ideal saturation may occur because of the finite ion energy distribution and cumulative ion-induced surface modification, but the dominant mechanism remains the selective removal of the oxidized surface layer generated during the modification step.

\section{Conclusion}

This work demonstrates the first plasma atomic layer etching process for diamond using cyclic oxygen plasma modification and krypton ion removal.

The O$_2$/Kr ALE process exhibits the characteristic self-limiting behavior of atomic layer etching and enables controlled material removal with sub-nanometer resolution. An etch depth of approximately \SI{6.85}{\angstrom} per cycle and a synergy of 53\% were achieved.

Surface analysis by SEM, AFM, and XPS shows that the etched diamond surfaces exhibit lower roughness \SI{1.1}{nm} than the pristine CVD \SI{1.23}{nm}, while the diamond surface and sub-surface bonding structure remain largely preserved, indicating essential structural damage-free.

These results establish plasma ALE as a powerful approach for precise, damage-controlled processing of diamond and provide a promising pathway toward advanced diamond electronic, photonic devices quantum sensing and quantum computing technologies.

\section*{Acknowledgements}

This work was carried out in 2022--2023 in the JFAST Laboratory, University of Tsukuba, Japan, and Institut N\'eel, CNRS, Universit\'e Grenoble Alpes, Grenoble, France.
This work was funded by IDEX ISP 2020-l'initiative d'excellence (Idex) Université Grenoble Alpes and supported by IDEX mobility grant.

\section*{CRediT Author Statement}

D.D. Tran: Writing -- review \& editing, Writing -- original draft, Methodology, Investigation, Formal analysis, Data curation, Conceptualization. C. Mannequin: Supervision, Methodology, Validation, Review. A. Traore: Process training, Scientific Discussion. M. Sasaki: Supervision. E. Gheeraert: Supervision, Methodology, Validation, Review.

\bibliographystyle{elsarticle-num}
\bibliography{references}

\end{document}